\documentstyle[12pt,a4,epsfig,here]{article}
\begin{document}
{\bf DESY 99-17}
\vspace*{1.2cm}

\begin{center}

{\Large \bf Registration of Atmospheric Neutrinos}  

{\Large \bf with the BAIKAL Neutrino Telescope NT-96}

\end{center}
\vspace{0.6cm}

V.A.Balkanov$^1$, I.A.Belolaptikov$^7$, L.B.Bezrukov$^1$, N.M.Budnev$^2$, 
A.G.Chensky$^2$,\\ I.A.Danilchenko$^1$, Zh.-A.M.Djilkibaev$^1$, 
G.V.Domogatsky$^1$, A.A.Doroshenko$^1$,\\
S.V.Fialkovsky$^4$, O.N.Gaponenko$^1$, 
A.A.Garus$^1$, T.I.Gress$^2$, A.M.Klabukov$^1$, A.I.Klimov$^6$, \\
S.I.Klimushin$^1$, A.P.Koshechkin$^1$, V.F.Kulepov$^4$, L.A.Kuzmichev$^3$, 
V.E.Kuznetzov$^1$,\\ S.V.Lovtzov$^2$, B.K.Lubsandorzhiev$^1$, M.B.Milenin$^4$,
R.R.Mirgazov$^2$, N.I.Moseiko$^3$,\\ V.A.Netikov$^1$, E.A.Osipova$^3$, 
A.I.Panfilov$^1$, Yu.V.Parfenov$^2$, A.A.Pavlov$^2$, E.N.Pliskovsky$^1$,\\ 
P.G.Pokhil$^1$, E.G.Popova$^3$, M.I.Rozanov$^5$, V.Yu.Rubtzov$^2$, 
I.A.Sokalski$^1$, Ch.Spiering$^8$,\\ O.Streicher$^8$, B.A.Tarashansky$^2$, 
T.Thon$^8$, R.V.Vasiljev$^1$, R.Wischnewski$^8$, I.V.Yashin$^3$

\bigskip

{\it

$^1$ Institute  for  Nuclear  Research,  Russian  Acad.  of   Sciences
(Moscow, Russia),\\
$^2$  Irkutsk State University (Irkutsk, Russia),\\
$^3$  Moscow State University (Moscow, Russia),\\
$^4$  Nizhni  Novgorod  State  Technical  University  (Nizhni   Novgorod,
Russia),\\
$^5$ St.Petersburg State  Marine  Technical  University  (St.Petersburg,
Russia),\\
$^6$ Kurchatov Institute (Moscow, Russia),\\
$^7$ Joint Institute for Nuclear Research (Dubna, Russia),\\
$^8$ DESY Zeuthen (Zeuthen, Germany)}

\vspace{0.9cm}

{\small
{\bf Abstract:} We present first neutrino induced events observed
with a deep underwater neutrino telescope. Data from  70 days
effective life time of the BAIKAL prototype telescope NT-96 
have been analyzed
with two different methods. With the standard track
reconstruction method, 9 clear upward muon candidates have been 
identified, in good agreement with 8.7 events expected from
Monte Carlo calculations for atmospheric neutrinos. 
The second analysis is tailored to 
muons coming from close to the opposite zenith. 
It yields 4 events, compared to 3.5 from Monte Carlo 
expectations. From this we derive a 90\%
upper flux limit of $1.1 \cdot 10^{-13}$ cm$^{-2}$ sec$^{-1}$
for muons in excess of those expected from atmospheric neutrinos 
with zenith angle $>$ 150 degrees and energy $>$ 10\,GeV.
}

\vspace{0.5cm}
\begin{center}
{\large \it submitted to Astroparticle Physics}
\end{center}
\newpage

\vspace*{1.5cm}

\section{Introduction}

The ultimate goal of large underwater neutrino telescopes is
the identification of extraterrestrial neutrinos of high
energies. The omnipresent background with respect to these neutrinos
are neutrinos generated by cosmic ray interactions in the
atmosphere of the Earth. While being a background with
respect to extraterrestrial neutrinos, 
atmospheric neutrinos can be used as a standard signal to
test and calibrate underwater neutrino detectors. 
Therefore, a primary challenge for these detectors is the
identification of upward muons generated in interactions
of atmospheric neutrinos. Taking into account that the flux
of downward muons at 1 km depth is about 6 orders of magnitude 
larger than the flux  of upward muons from
atmospheric neutrinos, 
this task is far from being trivial \cite{APP}.

Apart from being a test tool for underwater
telescopes, atmospheric neutrinos may indicate physics
beyond the standard model of particle physics. Actually,
from recent analyses of data from underground experiments,
evidence is found for oscillations of neutrinos 
\cite{Kamioka,Super-K,Macro,Soudan}. The search for such feeble
effects, however, requests a detailed understanding of the
detectors, high statistics and a relatively low
energy threshold.

The first part of this paper describes the method for full spatial
reconstruction of muon tracks  with the
four-string array NT-96 in Lake Baikal, and the criteria to reject
fake events from misreconstructed downward muons. We present 
results from the analysis of 70 days effective lifetime of NT-96,
taken in 1996.
 
In contrast to the standard reconstruction strategy, which supposes
$\ge$6 hits at $\ge$3 of the vertical strings, the second analysis
is performed also for events
with hits on less than 3 strings. This considerably increases
the effective area in vertical direction. Instead of beginning with
a reconstruction and then applying quality cuts, we start
with cuts effectively rejecting 
all events with the exception of nearly vertically upward moving
muons. Only for events surviving these cuts, a fit
of the zenith angle is performed.
An excess of such muons over the expectation value for
muons from atmospheric neutrinos could indicate neutrino 
production by  the annihilation of neutralinos -- the favored
super-symmetric candidate for cold  dark matter -- in the center of the
Earth.  By a method of that kind, two first neutrino events have been 
identified in NT-36, the very first small BAIKAL prototype array
\cite{NT-36}.
In the second part of this paper, we apply this strategy
to NT-96 data and derive
an upper limit on the excess muons from the center of the Earth.

\newpage

\section{The detector}

\begin{figure}[H]
\centering
\mbox{\epsfig{file=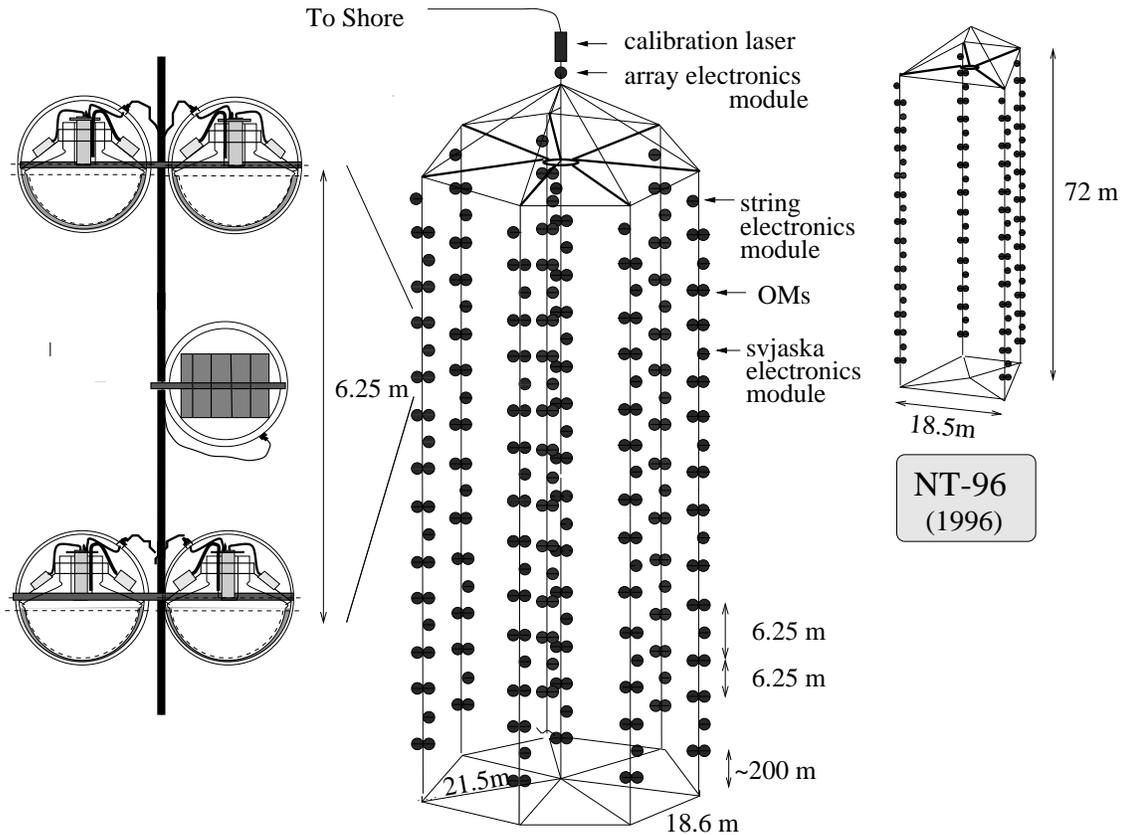,height=11cm}}
\caption[detector] {\small
Sketch of the Baikal Neutrino Telescope NT-200. Top right,
the 1996 stage NT-96 is shown.
}
\label{detector}
\end{figure}

The Baikal Neutrino Telescope \cite{APP,Project}
is operated in the Siberian Lake Baikal,
at a depth of 1.1\,km. In April 1998, the NT-200 array consisting
of 192 optical modules (OMs) at 8 strings was deployed 
(see fig.\ref{detector}).
The OMs are grouped in  pairs along the strings. They contain
37-cm {\it QUASAR} phototubes \cite{OM}. 
The two tubes of a pair are switched
in coincidence in order to suppress background from
bioluminescence and dark noise. A pair defines a {\it channel}.

A {\it muon trigger} is formed by the requirement of
$\ge N$ hit channels within 500 nsec.
$N$ is typically set to the value of 3 or 4. For such events,
amplitude and time of all fired channels are digitized and sent to shore.
The event record includes all hits within a time window
of -1.0 $\mu$sec to +0.8 $\mu$sec with respect to  
the muon trigger signal.

In April 1993, the first small prototype array, the detector
NT-36 with 36 OMs at 3 short strings, was launched and took 
data up to March 1995.  Results obtained with the 1993 version
of NT-36 as well as a detailed description of
detector, site and deployment procedure
have been published in \cite{APP}.
A 72-OM array, NT-72, run in 1995-96. In 1996, it was replaced by
the four-string array NT-96, which was replaced a year later
by a 144-OM detector. Since April 6, 1998, NT-200 with
192 OMs is taking data. 

NT-96 (48 channels) is shown at top right of fig.\ref{detector}. 
The OMs are fixed at four
strings each 72 m long. The strings are arranged at the
edges of a trapezoid with side lengths
of 3 $\times$ 18.5\,m, and 
1 $\times$ 10.2\,m. Each string consists of 12 pairs of OMs.
All OMs face downward, with the exception of the OMs of the
second and eleventh layers, which look upward. The distance
between downward oriented layers is 6.25\,m, 
the distance between layers facing to each other 
(layers 1/2 and 10/11) is
7.5\,m, the distance between back-to-back layers (2/3 and 11/12)
is 5.0\,m.

Results presented below are from 70.3 days data taking with NT-96,
starting at April 13, 1996.
During the first month, 47 out of all 48 
channels have been operating. In the following, nine channels failed, 
so that NT-96 
consisted of only 38 operating channels at the end of 
this period.

\section{Identification of Upward Muons by Full Reconstruction}

\subsection{The Reconstruction Procedure}

The reconstruction algorithm is based on the assumption that the
light radiated by the muons is emitted exactly under the  Cherenkov
angle (42 degrees) with respect to the muon path. The 
model of single minimum ionizing muon is a simplification,
since the direction of shower particles accompanying the muons 
is smeared around the muon direction. 
Atmospheric muons may occur
in bundles. Also, light may be  scattered on its way. In
Lake Baikal, however, the latter effect is negligible  since the
effective scattering length is much larger than the absorption length 
\cite{APP}.
The reconstruction procedure consists of the following steps:

\begin{enumerate}
\item A first quality analysis of the event which {\it a)}
   excludes events 
   far from being described by the model of a minimum ionizing muon, 
   and {\it b)} finds a first guess for the $\chi^2$ minimization.
\item Determination of the muon trajectory based on the 
   minimization of the function
\begin{eqnarray*}
\chi^2_t = \sum_{i=1}^{N_{hit}} (T_i(\theta, \phi, u_0, v_0, t_0)
    - t_i)^2 / \sigma_{ti}^2
\end{eqnarray*}

Here, $t_i$ are the measured times and $T_i$ the times expected for
a given set of track parameters. $N_{hit}$ is the number
of hit channels, $\sigma_{ti}$ are the timing errors. 
All  $\sigma_{ti}$ are set to 5\,nsec in the following procedures.
A set of
parameters defining a straight track is given by
$\theta$ and $\phi$ -- zenith and azimuth angle of the track,
respectively, $u_0$ and $v_0$ -- the two coordinates of
the track point closest to the center of the detector,
and $t_0$ -- the time the muon passes this point. 
For a fit allowing a three-dimensional reconstruction, $\ge$\,6 hits
on $\ge$\,3 strings are necessary (trigger {\it 6/3}).

\item  Rejection of most of the bad reconstructed events 
  with the help  of final quality  criteria.

\end{enumerate}

In the initial quality analysis (step 1),
an event has to pass the following criteria:

\vspace{-2mm}
\begin{itemize}
\item[a)] The time difference $\Delta t_{ij}$ for two channels
          $i$ and $j$ must obey the 
	following condition:
\vspace{-1mm}
\begin{eqnarray}
\label{eq:delta-t}
\mbox{$|\Delta t_{ij}|\cos\eta<r_{ij}/c+\delta$}, 
\end{eqnarray}      
\vspace{-1mm}

with $\eta$ being the 
	Cherenkov angle (42$^o$) and $r_{ij}$ the distance between
	channels. $\delta=$ 5\,nsec accounts for the timing error. 

\item[b)] For any two channels $i$ and $j$ on the same string, a
   region of zenith angles ${\theta^{min} -\theta^{max}}$ is determined 
   which is allowed  by the observed time differences $\Delta t_{ij}$:
\vspace{-1mm}
\begin{eqnarray}
\label{eq:thetalimit}
     {\cos({\theta^{min}+\eta})< \cos{{\eta}}
     \frac{c \, \Delta t_{ij}}{z_j-z_i}< \cos({\theta^{max}-\eta})}
\end{eqnarray}
\vspace{-1mm}

  Here ${z_i,z_j}$ are the $z$ coordinates of the channels.  
   If the regions of possible zenith angles for all pairs
   along a string do not overlap, the event is
   excluded.

\item[c)] Assuming the event is caused by a naked muon, 
   for every channel one can define a range of distances to
   the muon  which is consistent with the measured amplitude  of the
   channel. From this, for every channel {\it pair} the
   minimal ($\Delta t^{min}$) and maximal ($\Delta t^{max}$) allowed 
   time difference are defined, in dependence on the
   distance between the channels and the amplitudes.
   If for any pair the condition
   \mbox{$\Delta t^{min} <\Delta t^{exp}<\Delta t^{max}$}
   is violated, the event is rejected.
\end{itemize}
\vspace{-2mm}
Seventy percent of the events selected for reconstruction
(trigger {\it 6/3})
pass these criteria in both the experimental and the MC sample.
In the case of MC generated neutrino induced muons, the rate is
larger (80\%) due to the absence of muon bundles.

After the minimization (step 2) we apply final quality cuts
(step 3). Apart from the traditional cut
on $\chi^2_t$,
cuts on the following parameters are applied:
\begin{itemize}
\item the probability of non-fired channels not to be hit, $P_{nohit}$,
and fired channels to be hit, $P_{hit}$.
\vspace{-1mm}
\begin{eqnarray}
\label{eq:phit}
P_{hit} = (\prod^n_{i=1} p_i)^{1/n} \hspace{1cm}
P_{nohit} = (\prod^m_{j=1} p_i)^{1/m}
\end{eqnarray}  
\vspace{-1mm}
where $p_i$ is the hit probability for the $i$th of all $n$ fired
channels, and  $p_j$ the nohit probability for the $j$th of
all $m$ non-fired channels. 

\item the correlation function $A_{corr}$ of measured amplitudes
to the amplitudes expected: 
\begin{eqnarray}
\label{eq:ampcorr}
A_{corr} = \sum_{i=1}^{n} (A_i - \bar{A})(a_i - \bar{a})/
\sigma_a \sigma_A
\end{eqnarray}  
\vspace{-1mm}
with $A_i$ being the amplitude expected for channel $i$ on the
base on the fitted track hypothesis and a minimum ionizing
muon, $a_i$ the measured amplitude and $\sigma_a$ and $\sigma_A$
the corresponding dispersions.
\item an amplitude $\chi^2_A$ defined similar to the
time $\chi^2_t$ defined above.
\end{itemize}

\begin{figure}[H]
\centering
\mbox{\epsfig{file=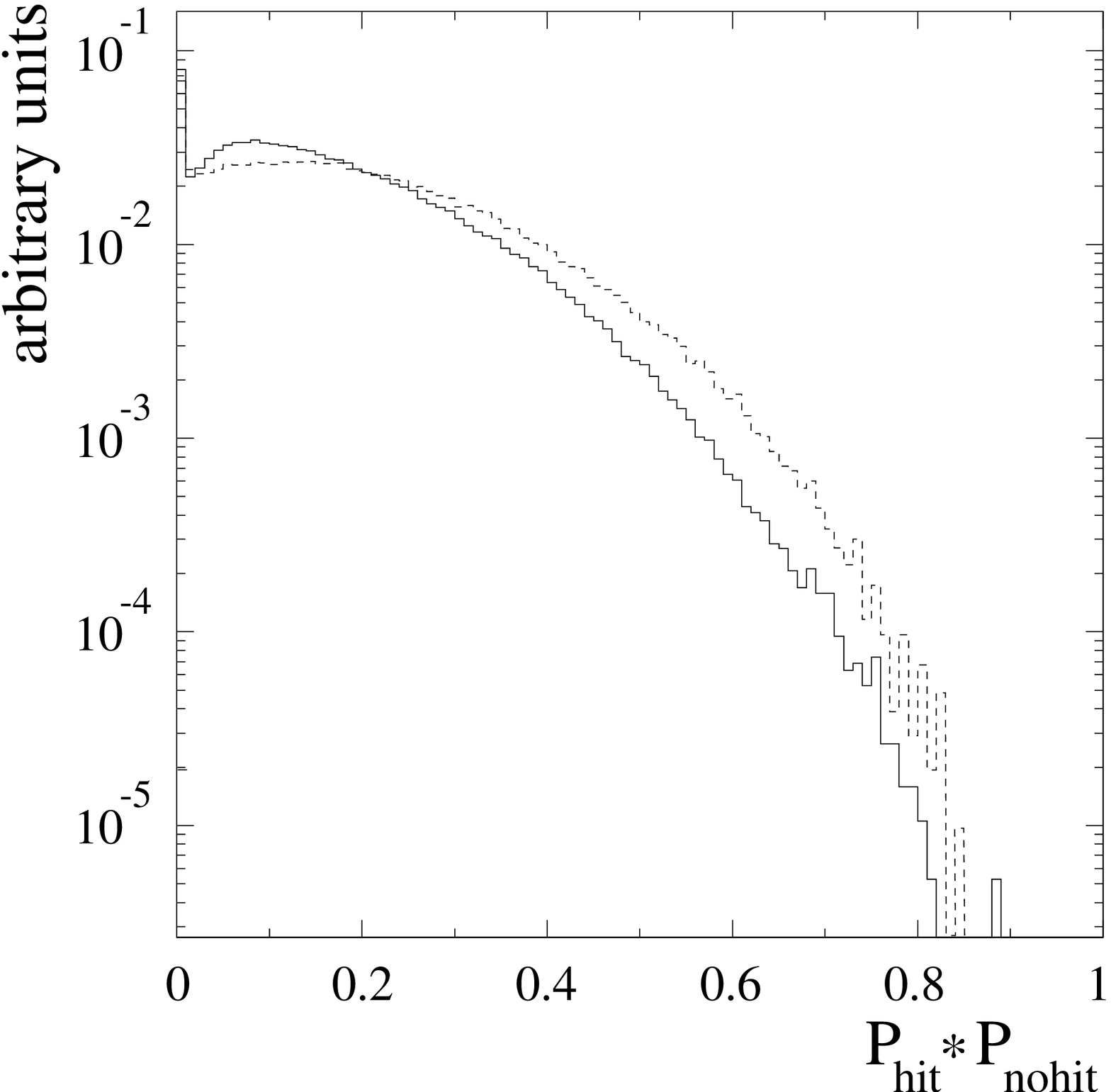,height=7.5cm,width=7.5cm}}
\mbox{\epsfig{file=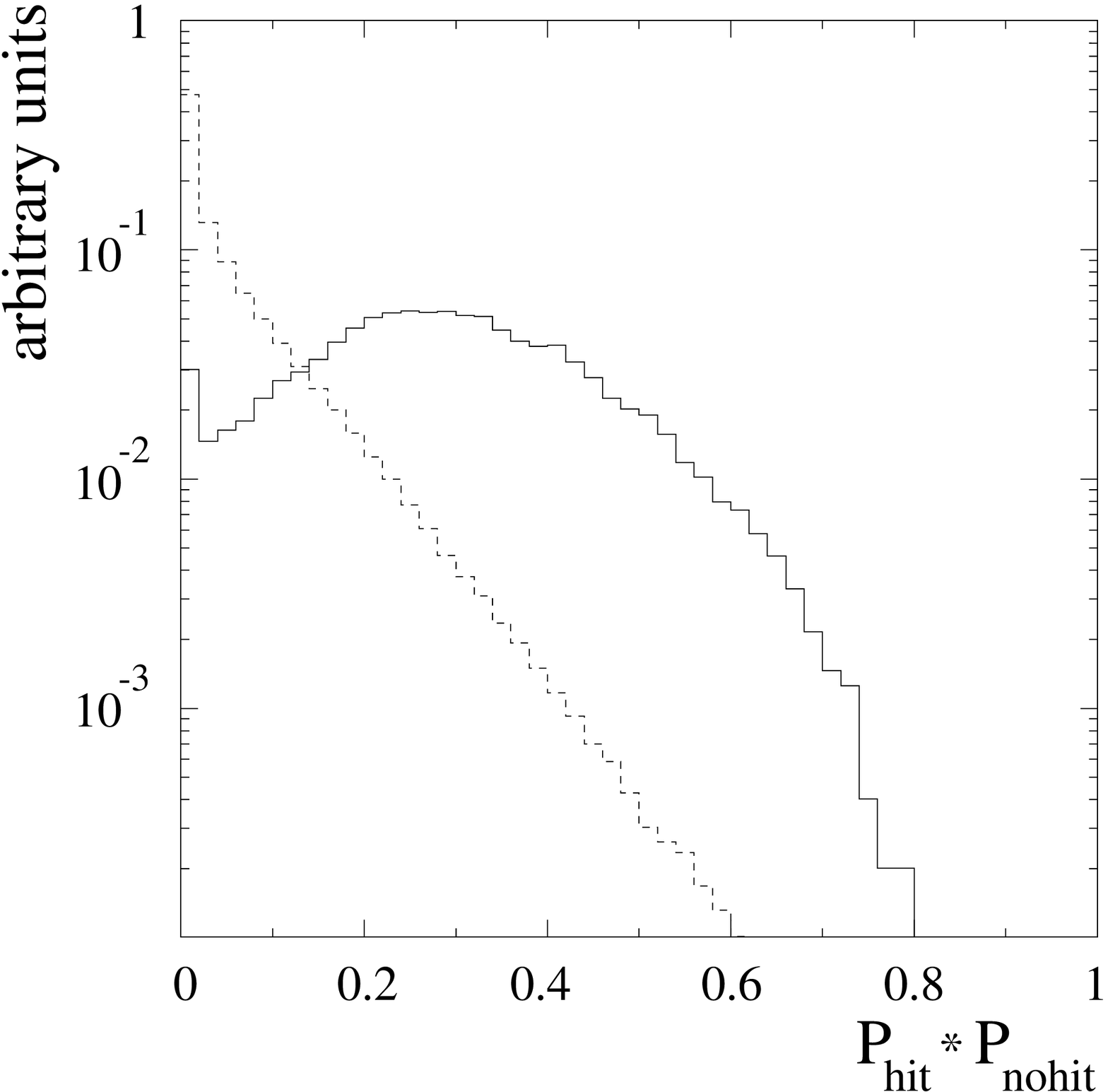,height=7.5cm,width=7.5cm}}
\caption[detector] {\small
Distribution of
$P_{hit} \cdot P_{nohit}$. Left: 
downward muon Monte Carlo events (dashed) and  
reconstructed experimental events (full line). Right:
MC fake events (dashed) and true MC upward muons (full line).
}
\label{parameters1}
\end{figure}

\begin{figure}[H]
\centering
\mbox{\epsfig{file=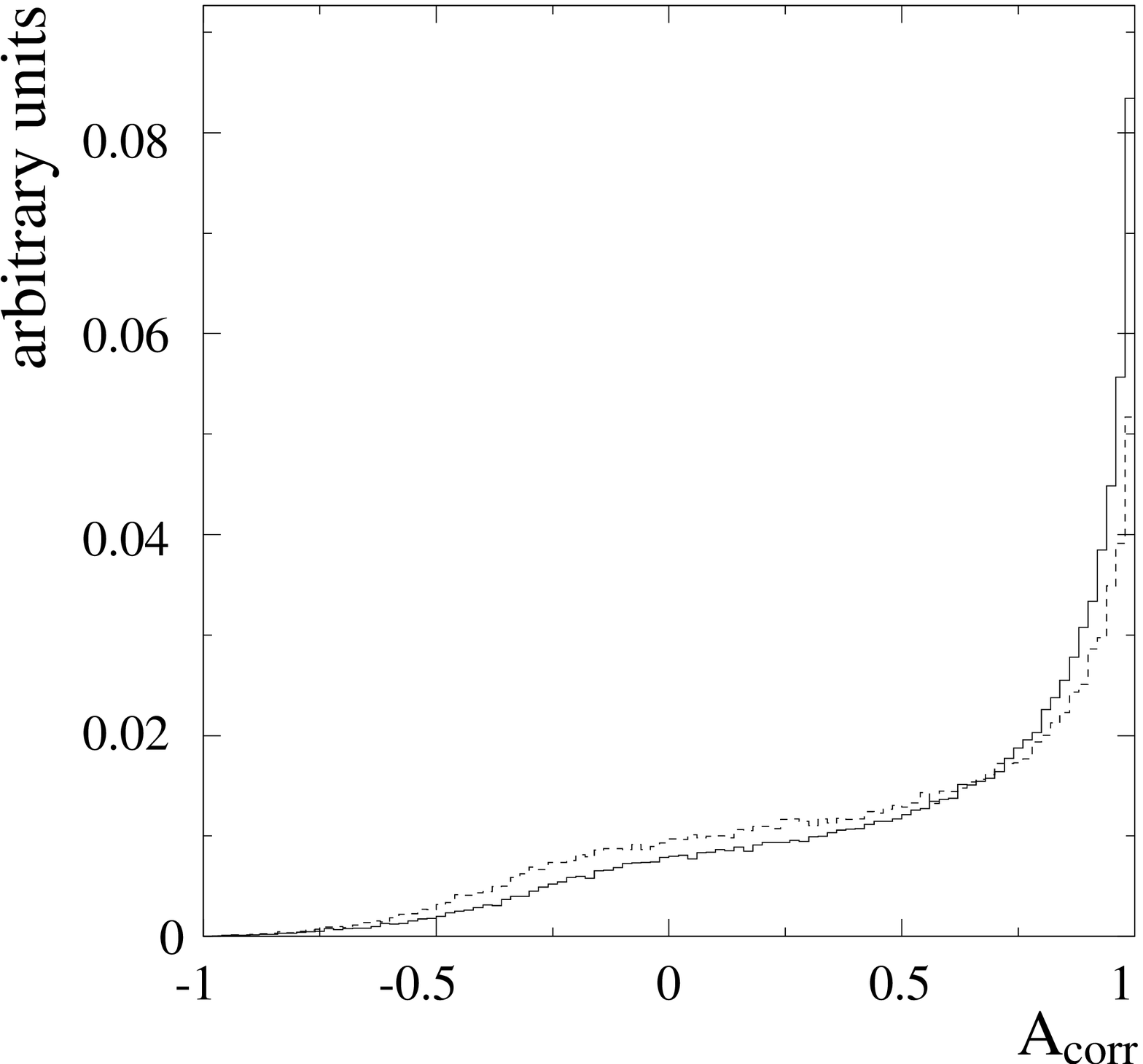,height=7.5cm,width=7.5cm}}
\mbox{\epsfig{file=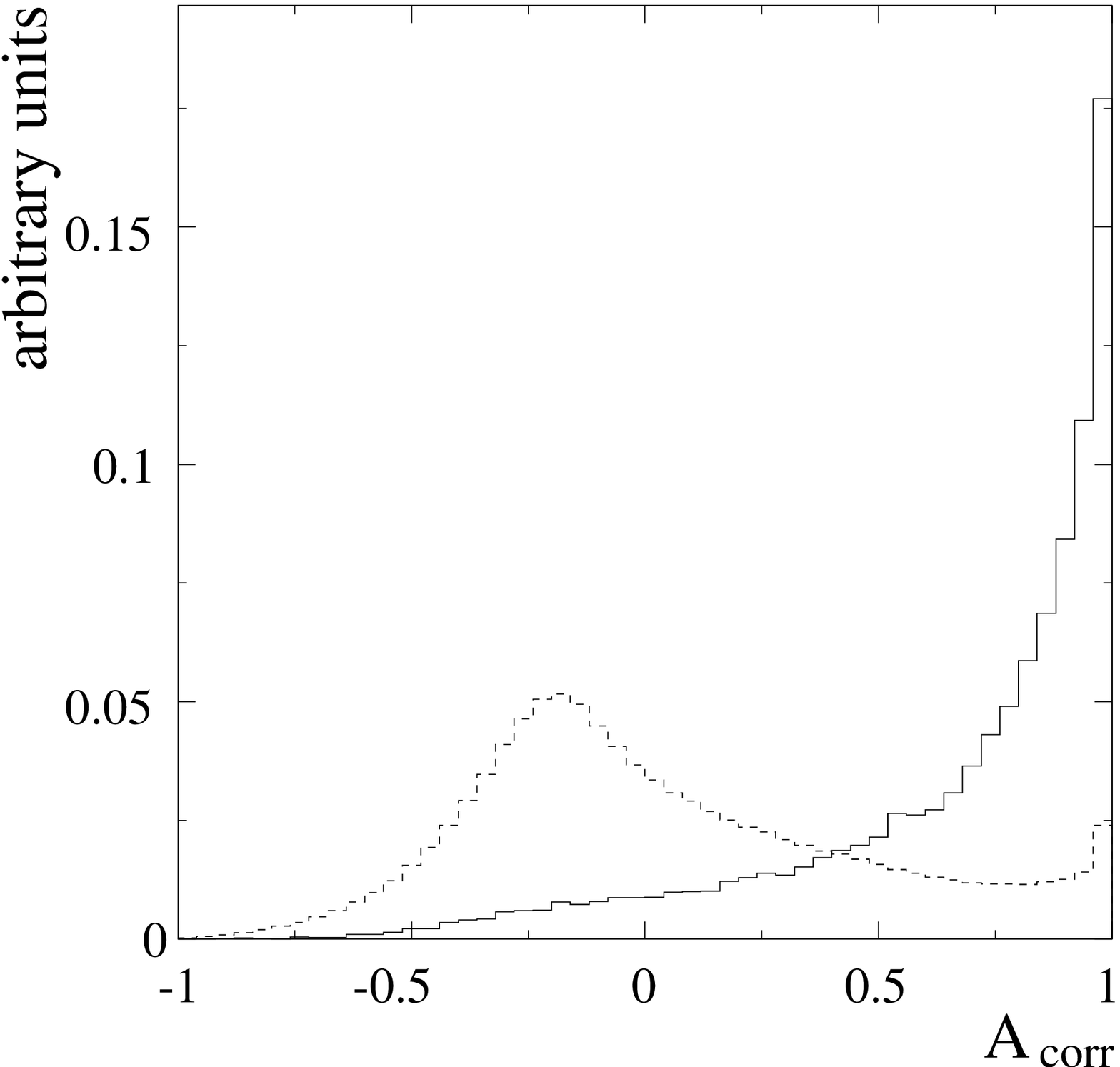,height=7.5cm,width=7.5cm}}
\caption[detector] {\small
Distribution of the amplitude correlation
$A_{corr}$.
Left: 
downward muon Monte Carlo events (dashed) and  
reconstructed experimental events (full line). Right:
MC fake events (dashed) and true MC upward muons (full line).
}
\label{parameters2}
\end{figure}

Figs.\ref{parameters1} and \ref{parameters2} show as an example 
the distributions of $P_{hit} \cdot P_{nohit}$ and the amplitude 
correlation $A_{corr}$ -- for downward muon Monte Carlo events  
compared to the total reconstructed experimental sample, and for
fake events compared to true upward muons (both MC). Atmospheric
muons have been generated according to \cite{Boziev} (see for
details \cite{APP} where also the Baikal detector simulation
is described), for atmospheric neutrinos the Volkova spectrum
\cite{Volkova} has been used.

To guarantee a minimum lever arm for track fitting,
we reject events with a projection of the most distant 
channels to the track ($z_{dist}$) smaller than a certain length. 
Due to the small transversal dimensions of {\it NT-96},
this cut excludes zenith angles close to the horizon, i.e.,
the effective area of the detector
with respect to atmospheric neutrinos is decreased considerably.

\begin{figure}[H]
\centering
\mbox{\epsfig{file=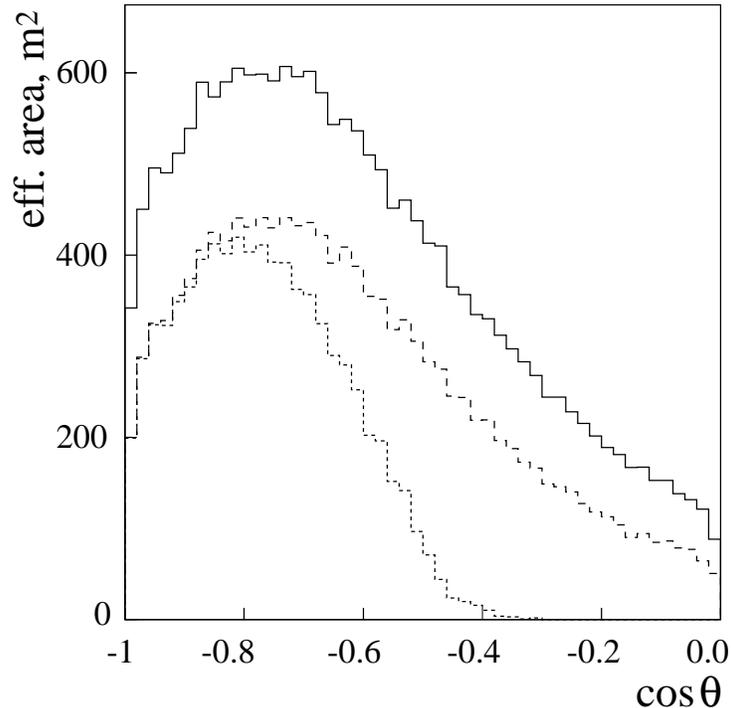,height=9.5cm}}
\caption[eff-area] {\small
Effective area for upward MC muons satisfying trigger {\it 9/3};
solid line - no quality cuts; dashed line~-~quality cuts 1-4;
dotted line - quality cuts 1-5. $\theta$ is the
MC generated zenith angle.
}
\label{eff-area}
\end{figure}

We have chosen the following 
cuts to separate well reconstructed
tracks:
\begin{itemize}
\item[{\bf 1:}] $\chi^2_t/NDF <$ 3.0
\item[{\bf 2:}] $P_{hit} \cdot P_{nohit} > $ 0.15
\item[{\bf 3:}] $\chi^2_A/NDF <$ 2.0
\item[{\bf 4:}] amplitude correlation $A_{corr} >$ 0.1
\item[{\bf 5:}] $z_{dist} >$ 35\,m
\end{itemize}
Fig.\ref{eff-area} shows the effective area for reconstructed upward
muons with $\ge$\,9 hits on $\ge$3\,strings, and for those passing
additionally cut 1-4 and 1-5, respectively, as function of
the zenith angle.

\newpage

\newpage

\subsection{Results Obtained with the Standard Analysis}

The efficiency of all criteria was tested using MC generated atmospheric
muons and upward muons due to atmospheric neutrinos. 
$ 1.8 \cdot 10^6$ events from
atmospheric muon events (trigger {\it 6/3}) have been simulated, 
with none of them
passing all cuts and being reconstructed as upward going muon.
Table 1 shows the number of events surviving various trigger/cut
levels (column 1) -- column 2 for the experimental data, column 3
for the Monte Carlo sample of downward muons, column 4
for the same sample normalized to 70.3 days, and column 5 for
upward muons from atmospheric neutrinos, again normalized to
70.3 days.

Rejection of all events with less than 9 hits 
results in a small decrease of the neutrino sensitivity (see
table 2), but reduces the background by a factor of 4.
This corresponds to 
the lowest curve in fig.\ref{eff-area}.

We have reconstructed $2.0 \cdot 10^7$ events
taken with {\it NT-96} from April to August 1996
(70.3 days). 
Nine events were reconstructed as upward going muons,
passed all quality cuts and triggered at least 9 channels
at 3 strings. This compares to 8.7 events expected from atmospheric
neutrinos.
Fig.5 displays one of the neutrino candidates. 
Top right the times of the hit channels are shown as
as function of the vertical position of the channel. 
At each string we observe the
time dependence, characteristically for upward moving particles.

\begin{center}
{\bf Table 1:} Number of events surviving various trigger/cut
levels 
\end{center}
\vspace{-9mm}
\begin{table}[h]
\begin{center}
\begin{tabular} {|l|c|c|c|c|} \hline
Trigger/cut & Experiment & MC atm $\mu$ & MC atm $\mu$ & MC atm $\nu$ \\
 & & & 70 days & 70 days \\ \hline \hline
6/3 &  $2.0 \cdot 10^7$ & $1.8 \cdot 10^6$ & $2.5 
\cdot 10^7$ & 51.6 \\ \hline
(1)+upward rec'd &
       $7.2 \cdot 10^5$ & $5.5 \cdot 10^4$ & $7.7 \cdot 10^5$ & 49.7 \\ \hline
(2)+quality cuts 1-4  & 5088 & 454 & 6361 & 17.3 \\ \hline
(3)+quality cut 5 & 30 & 0  &    - & 9.9 \\ \hline
(4)+9/3 & 9 & 0 & - & 8.7 \\ \hline
\end{tabular}
\end{center}
\end{table}

\vspace{0.3cm}

{\bf Table 2:} The fraction of events after the quality
cuts relative to the number of events surviving pre-criteria
and reconstruction for the triggers {\it 6/3} and {\it 9/3}
\vspace{-3mm}
\begin{table}[h]
\begin{center}
\begin{tabular} {|c|c|c|c|} \hline
\raisebox{-1.5ex}{Trigger cond.}&
\raisebox{-1.5ex}{Experiment}&
\raisebox{-1.5ex}{MC atm $\mu$}&
\raisebox{-1.5ex}{MC $\mu$ from $\nu$} \\ \hline \hline
6/3 &   0.19 &           0.21    &        0.20 \\ \hline
9/3 &   0.044&           0.056   &       0.175 \\ \hline
\end{tabular}
\end{center}
\end{table}


Applying eq.\ref{eq:thetalimit} 
not only to pairs at the same string, but
to all pairs of 
hit channels, one can construct an allowed region in both 
$\theta$ and $\phi$. For neutrino events this region
is requested to include regions below the
horizon.
This is demonstrated at the bottom right picture of fig.5.
The same holds for the other eight events, one of which is shown
in fig.\ref{sec-ambig}a). Fig.\ref{sec-ambig}b), 
in contrast, shows an ambiguous event not passing all quality cuts
and giving,
apart from the upward solution, also a downward solution. In this case
we assign the event to the downward sample.

The resulting angular distribution is presented in 
fig.\ref{ang_nu}a). Fig.\ref{ang_nu}b) shows the
measured angular distribution of the 9 neutrino
candidates weighted with the acceptance and compares it
to the spectrum of atmospheric neutrinos which was taken from
\cite{Volkova}. 
Table 3
gives the number of hit channels and of hit strings, and the zenith angle
for the nine neutrino candidates.
The average number 12 of hit channels agrees within errors with
the expected number of 11.2, the mean $\chi^2_t$ is 0.91, 
only one event has a $\chi^2_t$ value (2.45)
close to the cut limit of 3. 

\newpage

\begin{figure}[htb]
{\parbox[t]{65mm}{
\epsfig{file=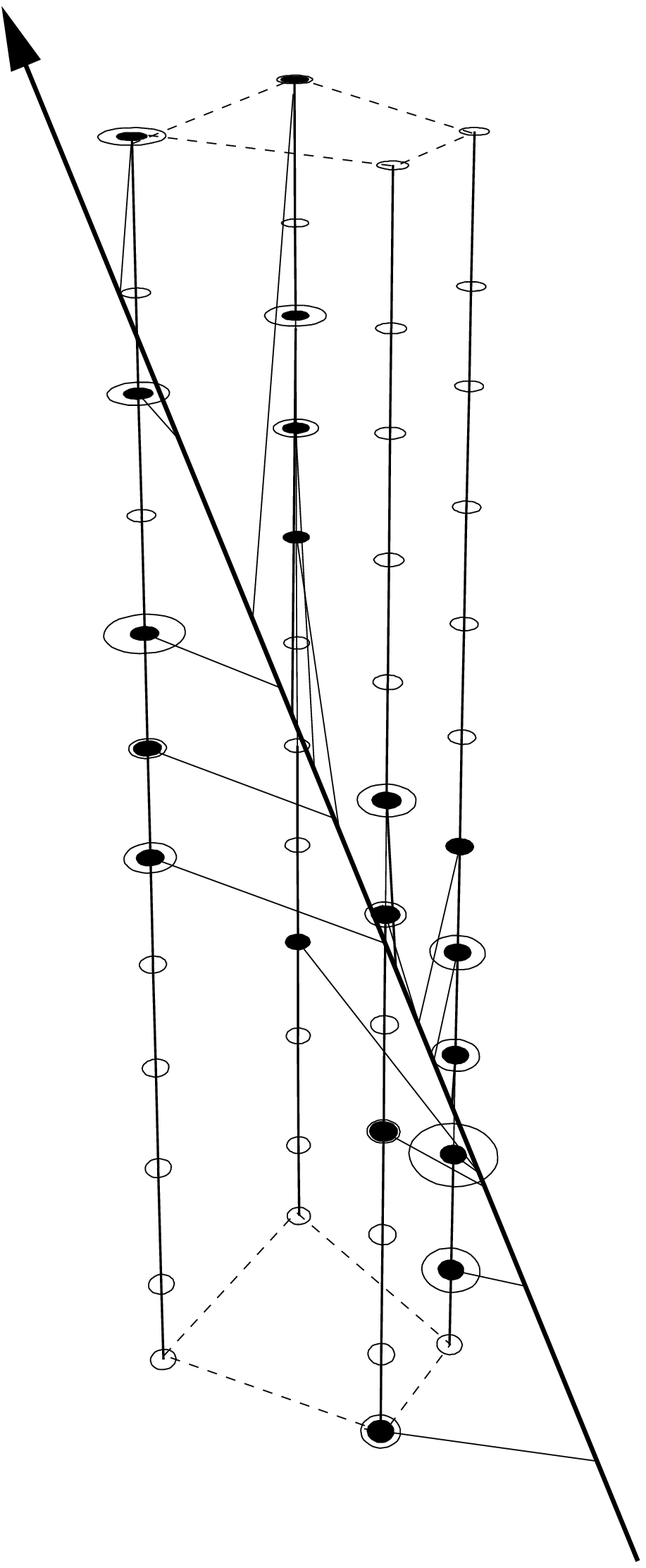,height=16cm}}}
\put(30,0){\parbox[t]{80mm}{
\epsfig{file=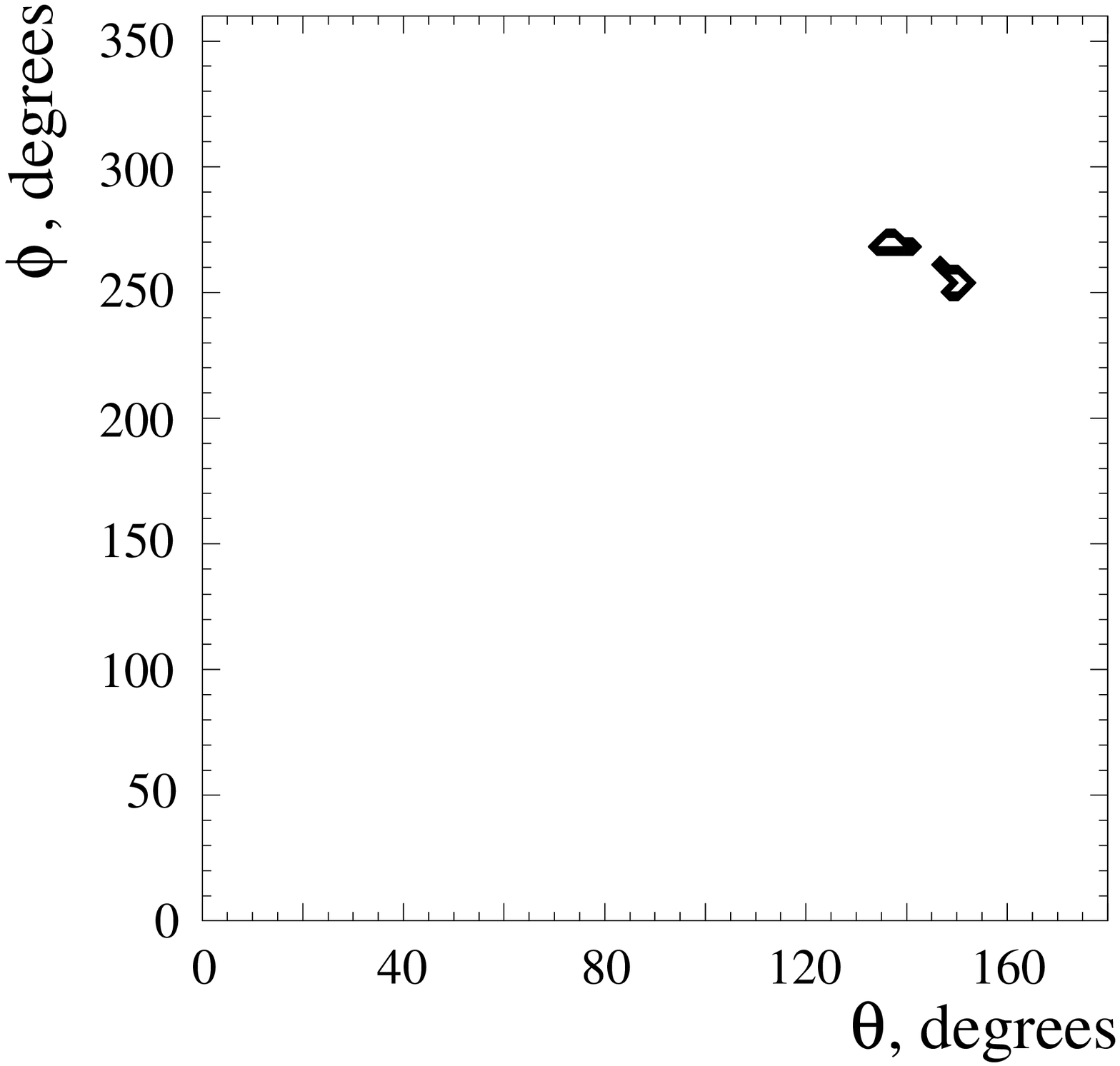,width=7.6cm}}
}
\put(30,230){\parbox[b]{80mm}{
\epsfig{file=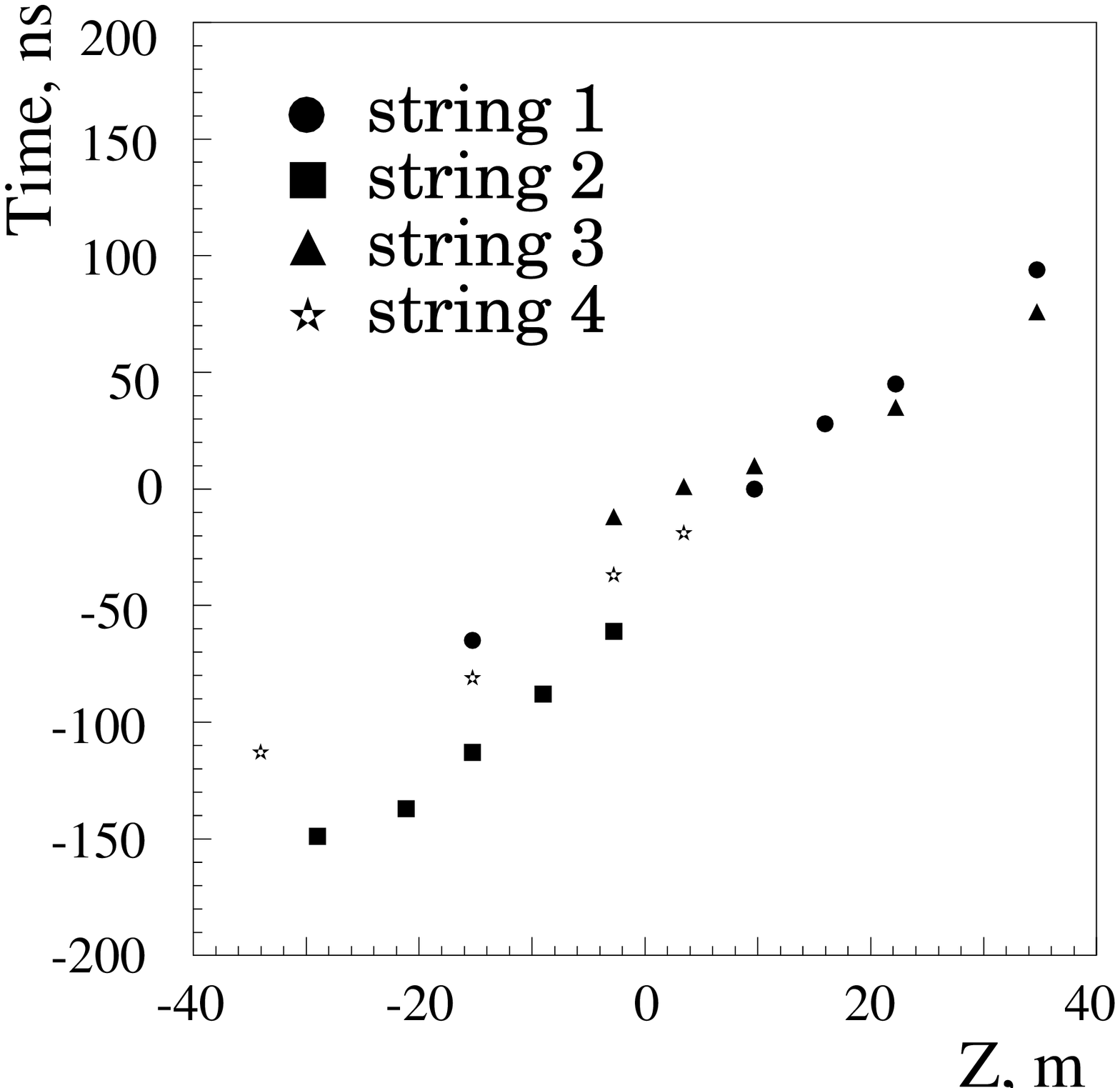,width=7.6cm}}
}
\caption[gold]{\small
A "gold plated" 19-hit neutrino event. {\it Left:} Event display.
Hit channels are in black. The thick line gives the 
reconstructed muon path, thin lines pointing to the channels mark the path
of the 
Cherenkov photons as given by the fit to the measured times. The sizes of 
the ellipses are proportional to the recorded amplitudes. {\it Top
  right:} Hit
times versus vertical channel positions. 
{\it Bottom right:}  The allowed $\theta/\phi$ regions (see text).
}
\end{figure}

\clearpage

\newpage

\begin{figure}[H]
\centering
\mbox{\epsfig{file=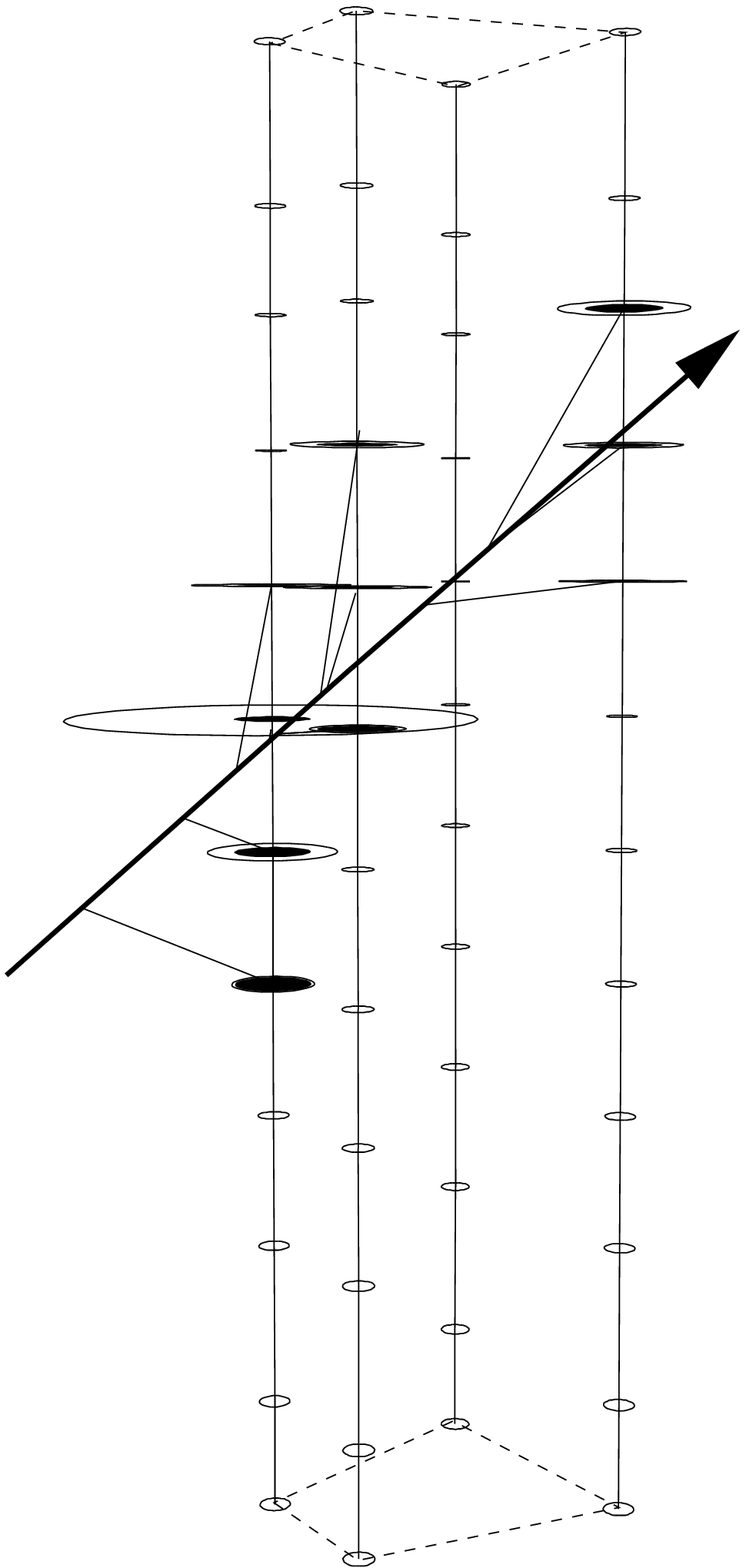,height=11cm}}
\mbox{\epsfig{file=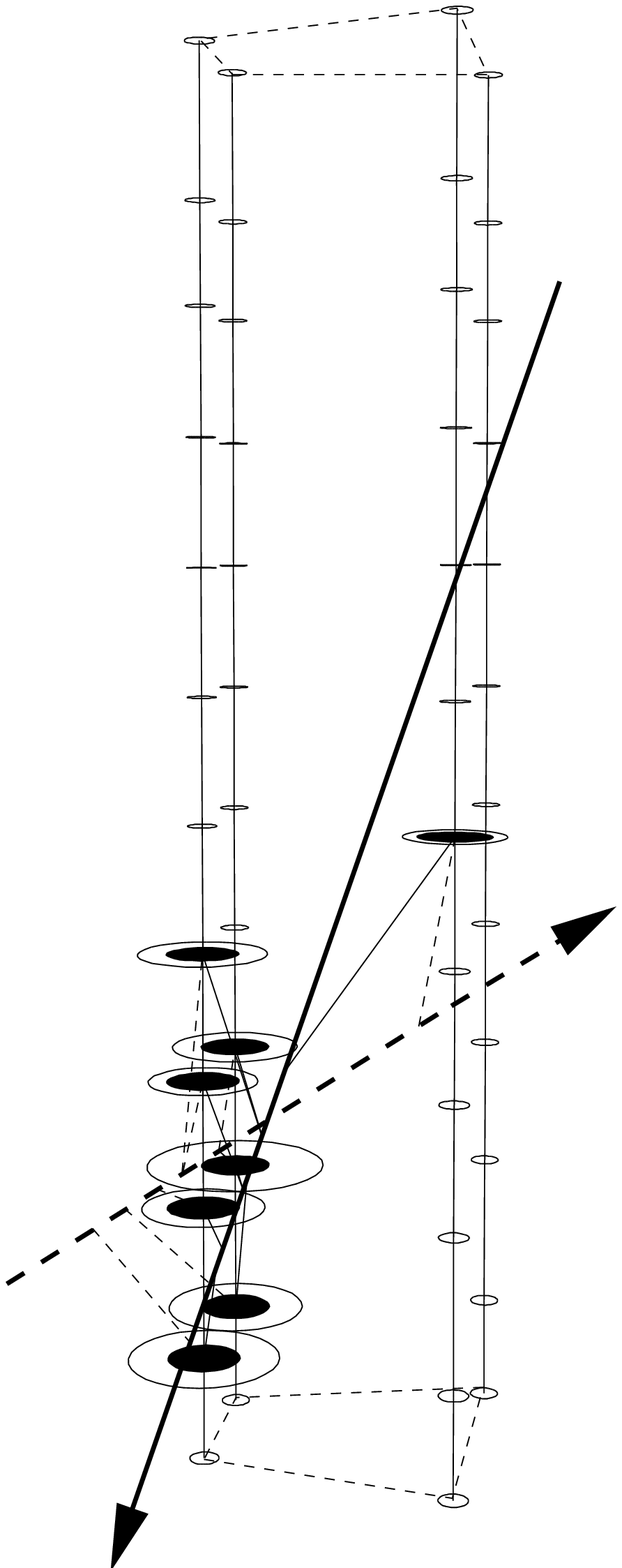,height=11cm}}
\caption[eff-area] {\small
{\it left:} An unambiguously reconstructed 10-hit neutrino candidate.
{\it right:} An ambiguous event reconstructed as a neutrino
event (dashed line) but not passing all quality cuts. This event
has a second solution above the horizon
(solid line) and was assigned to the sample of 
downward-going muons.
}
\label{sec-ambig}
\end{figure}                                  

\smallskip

\begin{center}
{\bf Table 3:} Parameters for the nine upward muon candidates
\end{center}

\smallskip
\begin{center}
\begin{tabular} {|c|c|c|c|} \hline
$\#$  & $N_{hit}$ & $N_{string}$ & $\theta$  \\ \hline \hline
{\it 1.} & 14 & 4 & 134.7$^o$  \\ \hline 
{\it 2.} & 19 & 4 & 152.3$^o$  \\ \hline 
{\it 3.} & 10 & 3 & 120.4$^o$  \\ \hline 
{\it 4.} & 16 & 4 & 145.5$^o$  \\ \hline 
{\it 5.} & 11 & 4 & 121.8$^o$  \\ \hline  
{\it 6.} &  9 & 4 & 138.6$^o$  \\ \hline  
{\it 7.} &  9 & 3 & 132.6$^o$  \\ \hline  
{\it 8.} & 11 & 4 & 136.9$^o$  \\ \hline  
{\it 9.} &  9 & 4 & 125.3$^o$  \\ \hline  \hline
\end{tabular}
\end{center}


\newpage

\begin{figure}[H]
\centering
\mbox{\epsfig{file=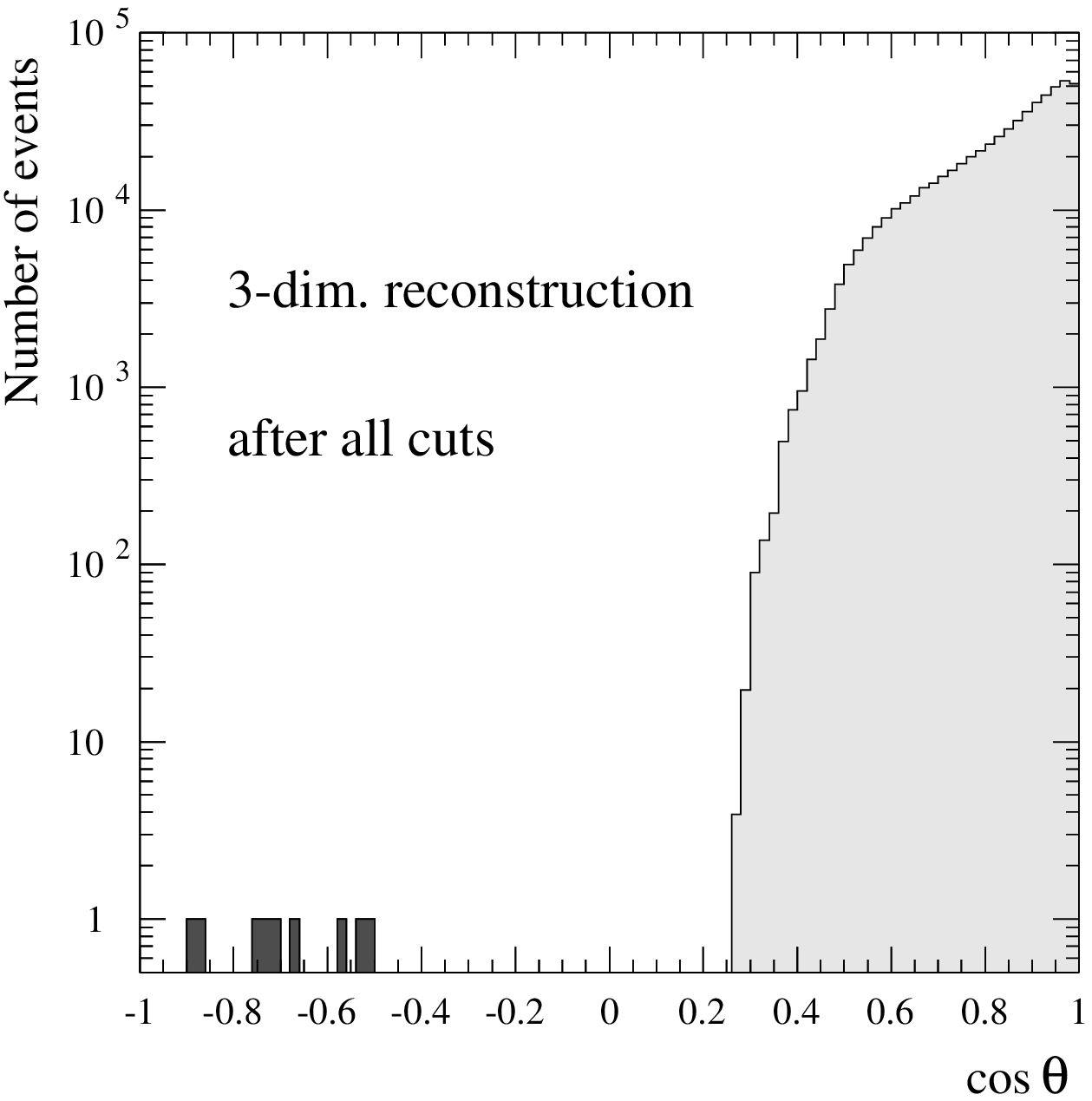,width=7.4cm}}
\mbox{\epsfig{file=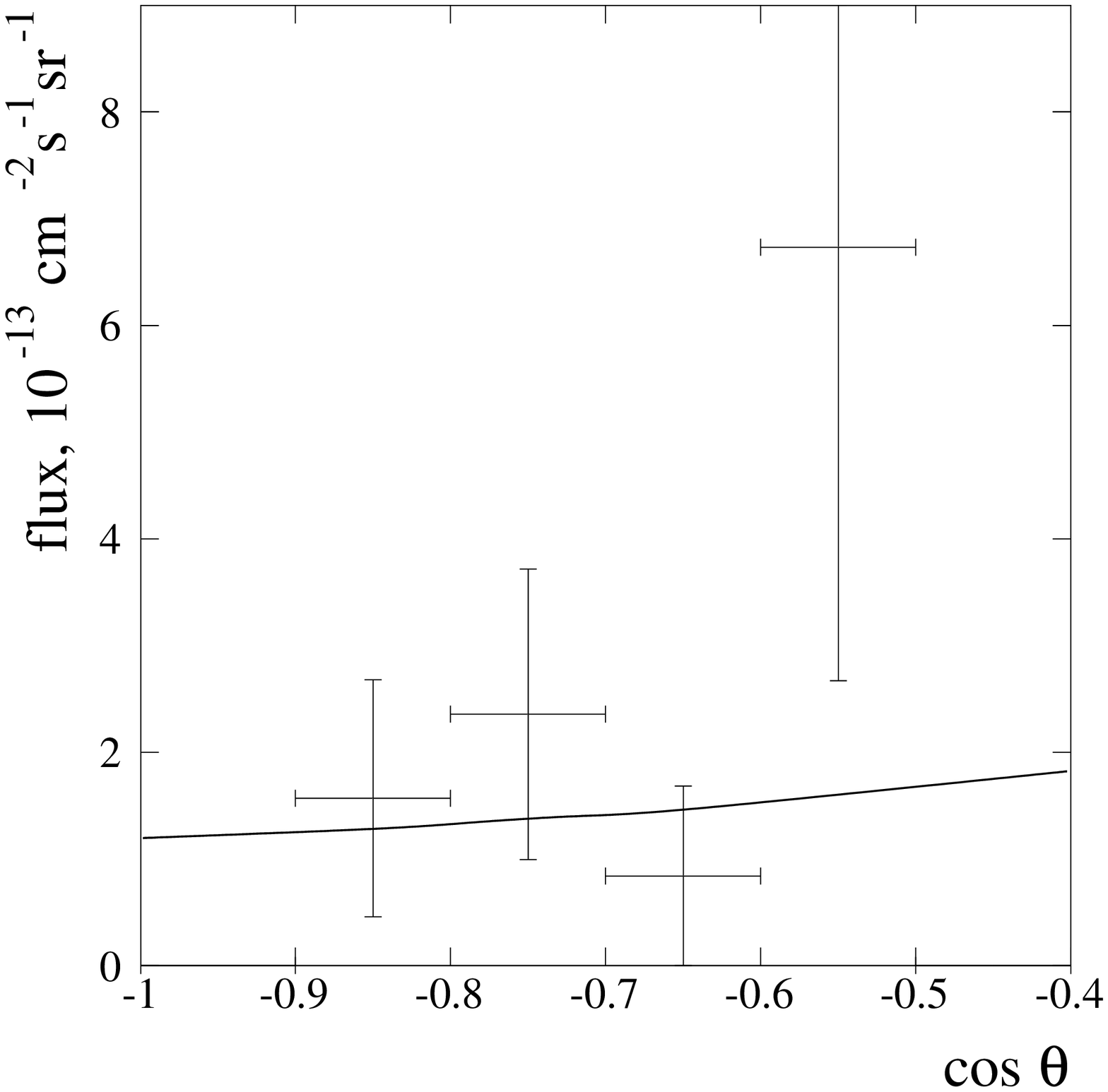,width=7.4cm}}
\caption[ang_nu] {\small
{\it left:} Experimental angular distribution of events satisfying trigger 
{\it 9/3}, all final quality cuts and the limit on $z_{dist}$ (see text).
{\it right:} The angular distribution below $\cos \theta$ = -0.4,
corrected for the acceptance. The
full line is obtained from the Volkova spectrum for atmospheric 
neutrinos \cite{Volkova}.
}
\label{ang_nu}
\end{figure}

\vspace{-5mm}

\section{Identification of Nearly Vertically Upward Moving Muons}    

\vspace{-3mm}

Different to the standard analysis, the
method presented
in this section relies on the application of
a series of cuts which are tailored to the response
of the telescope to nearly vertically upward moving muons.
The cuts remove events far away from the opposite
zenith including fake events. Fake events are 
mostly due misinterpreted
atmospheric muons with true zenith angles close to the horizon, and
to pair and bremsstrahlung showers below the array
which may give rise to upward traveling light fronts. 
The method does not aim a full reconstruction of all spatial
parameters. Therefore, also events with hits at only one or two 
strings are included in the analysis. The candidates
identified by the cuts are afterwards fitted in order to
determine the zenith angle.

\vspace{-2mm}

\subsection{Definition of the cuts}

\vspace{-2mm}

In order to optimize the cuts with respect to
the signal-to-noise
ratio, the following Monte Carlo samples have been
generated and analyzed:

\vspace{-3mm}

\begin{itemize}
\item Atmospheric neutrinos below horizon ($4.4 \cdot 10^3$ triggered 
events corresponding to 53 years data taking).

\item Minimum ionizing atmospheric
muons with zenith angles $>60^o$ ($1.5 \cdot 10^6$
triggered events corresponding to 3.6 years data taking).
\item showers generated by downward muons close to the array.
 Showers have been generated in a 
 $160 \times 160 \times 100 $m$^3$ box which includes the detector 
  center. The energies ranged from 10 GeV to 10 TeV and were
normalized to atmospheric muons (energy and angular
distribution, and production probability for
bremsstrahlung,
$3.7 \cdot 10^7$ triggered events corresponding to
 10 years data taking)

\end{itemize}

\vspace{-3mm}

We included all events with $\ge$4 hits along at least one
of the four strings. 

\vspace{-1mm}

\subsubsection{Timing}
\vspace{-2mm}
The first criterion cuts on the times measured along each of
those individual strings with more than 1 hit. It applies
to all combinations of hit channels along the string:

{\bf cut 1:} \hspace{1cm} $|(t_i - t_j) - \Delta z_{ij}/c| < 
a \cdot \Delta z_{ij} + 2\delta,$ \hspace{1cm} $(i<j)$.          

The $t_i, t_j$ are the arrival times at channels $i,j$, and
$\Delta z_{ij}$ is their vertical distance.
With $\delta$\,=\,5\,nsec accounting for the timing error,
the condition $ |(t_i - t_j) - z_{ij}/c| <  2\delta$ (i.e.
$a$=0) would cut for
a signal traveling vertically upward with the speed of light, $c$.

\begin{figure}[H]
\centering
\mbox{\epsfig{file=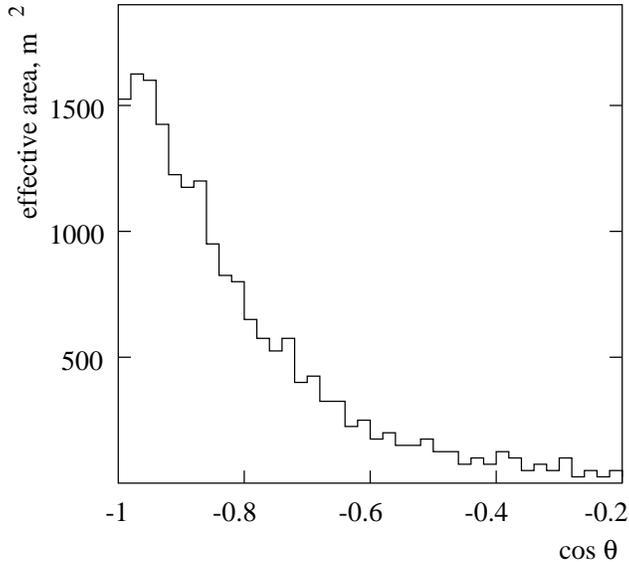,height=7.5cm}}
\vspace{-3mm}
\caption[timing] {\small
Effective area for muons from atmospheric neutrinos
passing cut 1.}
\label{timing}
\end{figure}

Setting $a$ to 1 nsec/m, the acceptance cone
around the opposite zenith is slightly increased.
Fig.\ref{timing} shows the effective area for muons from atmospheric
neutrinos as a function of $\cos\theta$ after cut {\it 1}.
All following cuts are applied to the sample of events
passing cut {\it 1}.

\subsubsection{Event Length}

\begin{figure}[H]
\centering
\mbox{\epsfig{file=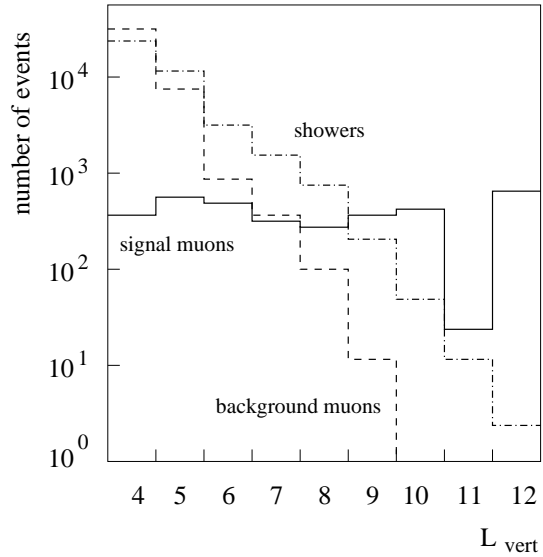,height=7.3cm}}
\caption[Lev] {\small
Distribution of the event length $L_{ev}$ for events from
atmospheric neutrinos (full line) as well as fake  events from
naked downward  muons (dashed line) and showers (dotted line),
respectively.  
}
\label{Lev}
\end{figure}

Events close to the vertical tend to be much more elongated
than hit patterns due to showers close to the array, and also than
fake downward muons, see fig.\ref{Lev}. The following cut
is the next  powerful after cut {\it 1} with respect to fake event rejection:

\medskip
{\bf cut 2:} \hspace{3cm} $L_{vert} = |i_{bot} - i_{top} + 1 | > 8$
\medskip


with $i_{top}$ and $i_{bot}$ being the numbers of the highest
and the lowest floor, respectively, which contain hit channels.
Numbers are counted from top to bottom. This cut defines a rather
sharp energy threshold at $E_{\mu} \approx$ 10\,GeV.

\subsubsection{Maximum Amplitude}

 
The distribution of the maximum amplitude in an event,
$A_{max}$,  differs strongly between atmospheric muons
(mean value 23 photoelectrons), naked-muon fake events and
shower events.

\smallskip
{\bf cut 3:} \hspace{4cm} $A_{max} < 50 $ photoelectrons,
\smallskip

accepts 90\,\% of the signal and removes about 
50\,\% of the shower events (for the parameter distributions
underlying this and the following cut values, see \cite{RussPrep}.


\subsubsection{Center of Gravity}

Naturally, events due to showers which emit light traveling
upward within the detector deposit most light in the lower 
part of the array. 
From the distribution of the center of gravity of hit channels for
signal and for shower events (not shown), we derive
a cut on the $z$ coordinate of the center of gravity:

\medskip
{\bf cut 4:} \hspace{3cm}
$COG_z = \sum^N_{i=1} (A_i \cdot z_i) / \sum^N_{i=1} (A_i) > 20$\,m,
\medskip

with $A_i$ being the amplitude of the $i$th channel and $N$ the 
number of all hit channels.



\subsubsection{Causality Condition}

The timing cut {\it 1} refers to the time differences between hit channels
along one string. However, one can also relate the hits
at different strings to each other assuming the particle
is moving nearly straight upward. For this reason, we define
a parameter

\begin{displaymath}
t_{basic,i} = t_{bot} + \frac{t_{top} - t_{bot}}{z_{top}-z_{bot}} 
\cdot (z_i - z_{bot})
\end{displaymath}

Here, $z_{bot}$ and $z_{top}$ are the $z$ coordinates of the lowest
and the highest fired channel at 
a string  with $N_{hit} \ge 4$,
respectively, and 
$t_{bot}$ and $t_{top}$  are the corresponding times.
$z_i$ is the $z$ coordinate of a hit channel at another string.
In the case of a straight upward muon, the time $t_i$
measured for this channel should be close to $t_{basic,i}$.

From the distribution of the largest values
of $|t_i - t_{basic,i}|$ found for each event, we derive
the following "causality cut": 

\medskip
{\bf cut 5:} \hspace{3cm} 
$\tilde{t} =$ max$(|t_i - t_{basic,i}|) < 60$ nsec.
\medskip


\subsubsection{Minimum time difference between top and bottom channels}             

According to cut {\it 2}, ($L_{vert}>8$), the distance between
{\it top} and {\it bottom} hit channels should be at least
50 m. This corresponds
to a travel time of 166 nsec for a relativistic muon. 
However, the actual difference in arrival times
may be slightly lower due to light scattering, shower particles
accompanying the muon and experimental errors.
We request that for all combinations of top and bottom hits
from various strings

{\bf cut 6:} \hspace{2cm} 
      $t_{tot} = \mbox{min}(t_{top,i} - t_{bot,j}) >$ 150 nsec

where $i(j)$ denotes the string number.



\newpage

\subsection{Application to Experimental Data}

Within 70 days of effective data taking, $8.4 \cdot 10^7$ events
with the muon trigger $N_{hit} \ge 4$ have been taken. 8608
of them have $\ge$4 hits at one string and pass cut {\it 1}.
Table 4 
summarizes the number of events
from all 3 events samples (MC signal and background, and experiment)
which survive the subsequent cuts. The experimental sample
consists of $8.4 \cdot 10^7$ events taken within 70.3 days
effective live time with the hardware trigger $N_{hit} \ge 4$.
The Monte Carlo samples are normalized to the same time interval.

\vspace{3mm}
{\bf Table 4:} The expected number of atmospheric neutrino events 
and background
events, and the observed number of events after cuts {\it 1--6}.

\begin{center}
\begin{tabular} {|c|c|c|c|c|c|c|} \hline
after cut {\cal N}$^o$  $\rightarrow$ & 1 & 2 & 3 & 4 & 5 & 6 \\ \hline \hline
atm. $\nu$, MC & 11.2 & 5.5 & 4.9 & 4.1 & 4.8 & 3.5 \\ \hline
background, MC & 7106 & 56 & 41 & 16 & 1.1 & 0.2 \\ \hline
experiment & 8608 & 87 & 66 & 28 & 5 & 4 \\ \hline \hline
\end{tabular}
\end{center}

\vspace{2mm}

The disagreement between background Monte Carlo and experiment 
in the first columns is mostly due to the fact that the Monte-Carlo 
was simplified in order to save CPU time and does not include 
low energy showers ($E_{shower}<$\,10\,GeV) accompanying the muon.
Whereas the number of MC background events continues to fall
with additional cuts, experimental data and  MC signal events
apparently reach a kind of asymptotic value which is only
slightly affected by the last cut. The parameters of the four
experimentally selected events are given in table 5 
together with the results of the zenith angle fit.
We note that all parameters values fall in the most
probable range for signal events which had passed the time cut.
Moreover, they are not close to the cut limits, i.e.,
a reasonable variation of the cuts would not change
the number of events passing the cuts. We conclude that
we have separated 4 atmospheric neutrino candidates with
expectation values of 3.5 for the signal itself and 0.2 for
the background from fake events. 

\vspace{3mm}
\hspace*{5mm}{\bf Table 5:} Fit results for the four 
upward muon candidates surviving
all cuts {\it 1--6}

\begin{center}
\begin{tabular} {|c|c|c|c|c|c|c|c|c|c|} \hline
$\#$  & $N_{hit}$ & $N_{string}$ & $L_{ev}$ & $A_{max}$ & 
$COG_{z}$ & $\tilde{t}$ & $t_{tot}$ &  $\theta$ & $\chi^2$/NDF \\ 
 &  &  &  & (p.e.) & (m) & (nsec) & (nsec) & (degree) &  \\ \hline
 & 19 & 4 & 12 & 12.5 & 34.8 &  0 & 190 & 152 & 0.44 \\ \hline
 &  9 & 2 & 12 & 21.0 & 23.5 & 38 & 270 & 161 & 0.05 \\ \hline
 &  9 & 2 & 10 & 13.0 & 26.0 & 22 & 202 & 162 & 0.14 \\ \hline
 &  7 & 2 & 10 &  4.0 & 32.0 & 23 & 188 & 170 & 0.39 \\ \hline \hline
\end{tabular}
\end{center}

\begin{figure}[H]
\centering
\mbox{\epsfig{file=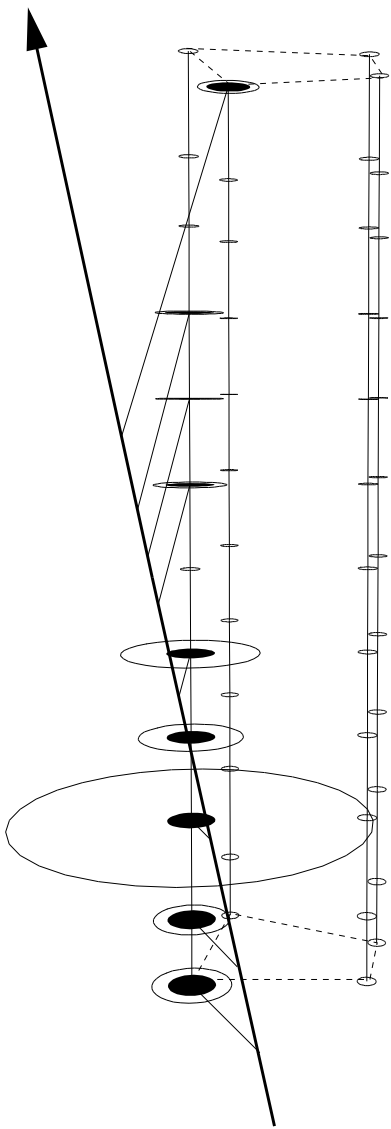,height=15cm}}
\caption[Lev] {\small
The third event of table 4,
with 9 hits at  2 strings
and 161$^o$ reconstructed zenith angle.
}
\label{vert2}
\end{figure}
\begin{figure}[H]
\centering
\mbox{\epsfig{file=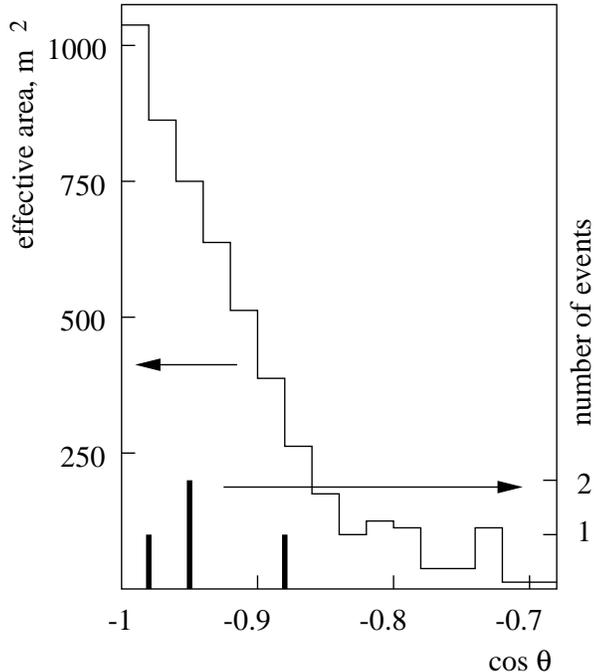,height=9cm}}
\caption[Lev] {\small
The effective area of {\it NT-96} after cuts {\it 1-6} (histogram) compared
to the reconstructed angles of the four events passing these cuts.
}
\label{angdis-vert}
\end{figure}

The first of the four events is the gold plated 19-hit
event which was also found in the full-reconstruction analysis.
Fig.\ref{vert2} shows the second event with 9 hits
at 2 strings. Fig.\ref{angdis-vert} gives the angular distribution of
the four events compared to the effective area after all cuts.
With a nearly flat distribution for atmospheric
neutrinos over the small $\cos \theta$ range considered, the
observed events should be distributed according to the
effective area. In fact, the distribution of the four observed events
is compatible with this assumption.

\vspace{10mm}

\subsection{Limit on the Flux of Muons from the Center of the Earth}

One class of candidate particles for cold dark matter has an
interaction strength typically for the weak scale. Weakly
Interacting Massive Particles (WIMPs) would have been produced
during the big bang and are diffusing through the universe.
They are expected to be scattered off coherently by nuclei,
e.g. in the Sun or the Earth. Loosing energy they eventually
fall below the escape velocity and are gravitationally trapped
at the center of these bodies. Once accumulated, annihilation
processes set on. The WIMP density builds up
until equilibrium is reached. The leptonic decays due to annihilation
into heavy quarks or W-pairs yield neutrinos of 
energies in the GeV-to-TeV range,
depending on the WIMP mass. The most favoured candidate for
WIMPs is the neutralino. In the Minimal
Supersymmetric Standard Model (MSSM), neutralinos
are linear combinations of the neutral gauginos
$\tilde{B}, \tilde{W}_3$, and of the neutral higgsinos $\tilde{H}_1^o, 
\tilde{H}_2^o$.
The lightest of these combinations
would be the stable candidate for cold dark matter.
The neutralino capture rate in celestial bodies, the
annihilation cross sections, the fragmentation functions into fermions,
gauge bosons and Higgs particles depend on the various parameters
of the MSSM (see \cite{Bottino,Berg} and references therein).

Underground detectors like {\it Baksan}, MACRO and {\it Kamiokande}
set upper limits on the flux of muons from the center of the
Earth of a few times $10^{14}$\,cm$^{-2}$\,sec$^{-1}$
\cite{Boliev,Montaruli,Mori}. These limits already exclude a certain
region of the SMMS parameter space. Neutralino masses below
25 GeV are excluded by LEP.

Table 5 shows the number of events detected in NT-96 within four zenith
angle regions about the opposite zenith, and compares it to
expectations from atmospheric neutrinos.
From the non-observation of a significant excess, the flux limits given
in the third column are obtained. They are compared to the limits 
given by {\it Baksan}, MACRO and {\it Kamiokande} in the last 3 columns.
In spite of the effective area of NT-96 for nearly vertically muons being
about 1000 m$^2$ (see fig.\ref{angdis-vert}), the
limits obtained are 4-7 times wore than the best published values,
since the only 70 days have been analyzed until now. However,  the
result illustrates the potential of underwater experiments with
respect to the search for muons due to neutralino annihilation.

\bigskip

{\bf Table 5:} The number of events detected ("Data") and expected  
from atmospheric neutrinos ("Bg") for {\it NT-96}, as well as
the $90\%$ C.L. upper limits on the muon flux from the center of the Earth
for four regions of zenith angles obtained {\it NT-96, Baksan}, MACRO and
{\it Kamiokande}.

\smallskip
\begin{center}
\begin{tabular}{|c|c|c|c|c|c|c|} \hline
        &  &    &\multicolumn{4}{c|}{Flux limit       ($10^{-14}
         \, \cdot $  cm$^{-2}$ \, sec$^{-1})$} \\ \cline{4-7} 
Zenith   & Data & Bg  & {\it NT-96}  & {\it Baksan} & {\it MACRO}  & {\it Kamiokande}  \\ 
angles   &      &      &$>10GeV$       &$>1GeV$        &$>1.5GeV$      &$>3GeV$ \\ \hline
$\geq 150^{\circ}$  & 4      & 3.7   & $11.0$      & $2.1$ & $2.67$ &$4.0$ \\ \hline
$\geq 155^{\circ}$  & 3      & 2.6   & $9.3 $      & $3.2$ & $2.14$ &$4.8$ \\ \hline
$\geq 160^{\circ}$  & 2-3    & 2.3   & $ 5.9-7.7 $ & $2.4$ & $1.72$ &$3.4$ \\ \hline
$\geq 165^{\circ}$  & 1      & 1.3   & $4.8$       & $1.6$ & $1.44$ &$3.3$ \\ \hline
	\end{tabular} 
\end{center}

\bigskip

\section{Conclusions}

Twelve  neutrino
candidates have been separated
from 70 days effective lifetime of the four-string 
detector {\it NT-96} in Lake Baikal. 
Nine of them have hits at $\ge$ 3 strings and
are fully reconstructed (zenith and azimuth angle). 
The other three events have hits at 2 
strings but still yield an unambiguous zenith reconstruction. The observed
number as well as the angular distribution coincide with
expectations from atmospheric neutrinos.

This is the first successful {\sl experimentum crucis} for
the operation of underwater neutrino telescopes.
From the agreement with expectations, we derive an
upper flux limit of $1.1 \cdot 10^{-13}$ cm$^{-2}$ sec$^{-1}$
for muons with zenith angle $ > 150^o$ and 
$E_{\mu} >$10\,GeV in excess of those expected from atmospheric neutrinos. 
Such muons might have been produced in neutralino
annihilations in the center of the Earth.

At present, the {\it NT-200} telescope
is taking data. It contains twice the 
number of optical modules as {\it NT-96}.
From the fully operational {\it NT-200} array, about one neutrino event per
day is going to be separated. {\it NT-200} will be primarily
used to study atmospheric neutrinos and to search for upward muons 
from WIMP annihilation.

\end{document}